\documentclass[11pt]{article}
\usepackage{amsfonts}
\usepackage{pstricks}
\usepackage{pst-node}
\usepackage{epsfig}
\usepackage{epsfig}

\begin{document}

\def\ba{\begin{eqnarray}}
\def\ea{\end{eqnarray}}

\begin{titlepage}
\title{\bf The Quadratic Symmetric Teleparallel Gravity in Two-Dimensions}
\author{ Muzaffer Adak  \\
 {\small Department of Physics, Pamukkale University}\\ {\small 20017 Denizli, Turkey} \\
 {\small \tt madak@pau.edu.tr} \\
 \\      Tekin Dereli  \\
 {\small Department of Physics, Ko\c{c} University} \\
 {\small 34450 Sar{\i}yer-Istanbul, Turkey} \\
 {\small \tt tdereli@ku.edu.tr} }
  \vskip 1cm
\date{ }
\maketitle

 \thispagestyle{empty}

\begin{abstract}
 \noindent
A 2D symmetric teleparallel gravity model is given  by a generic
4-parameter action that is quadratic in the non-metricity tensor.
Variational field equations are derived.  A class of conformally
flat solutions are found. We also give   static or cosmological
solutions that need not be in this class.
\end{abstract}
\end{titlepage}

\section{Introduction}

In order to investigate the physical implications of  General
Relativity (GR) some simple limiting cases are often sought. The
low energy, static limit is attained by assuming the existence of
a time-like Killing vector. However, such configurations have no
dynamics. Symmetric limits at arbitrary energy scales on the other
hand are obtained by assuming one or more space-like Killing
vectors. For example, the assumption of spherical symmetry reduces
the gravitational action upon integrating over the angular
variables to an effective 2D gravity model in $(t,r)$ coordinates.
Such 2D gravity models have recently been used much to discuss
black hole dynamics, quantized gravity or numerical relativity
\cite{grumiller}. It is well known that Einsteinean gravity in 2D
has no dynamical degrees of freedom. One way of making this theory
dynamical is to couple a dilaton scalar. It is also possible to
introduce further dynamical degrees of freedom by going over to a
non-Riemannian space-time by the inclusion of torsion and
non-metricity. Generalized 2D gravity models with curvature and
torsion or teleparallel theories with only torsion have been
considered before (See the references in  \cite{obukhov2004,
adak2006b}). On the other hand  models including  non-metricity
received less attention \cite{dereli1994, hehl1995}. The symmetric
teleparallel gravity (STPG) with vanishing curvature and torsion
in four-dimensions has been introduced relatively recently
\cite{nester1999, adak2005, adak2006}.  Here we consider a 2D
symmetric teleparallel gravity model given by the most general
4-parameter action that is quadratic in the non-metricity tensor.
We give a class of conformally flat solutions to the variational
field equations obtained by a re-scaling of the 2D Minkowski
metric.  We also discuss static and cosmological solutions that
need not be in this class.

\section{ Mathematical Preliminaries}

The triple $ \{M,g,\nabla \} $ denotes the space-time where $M$ is a
$2$-dimensional differentiable manifold, $ g $ is a non-degenerate
metric and $ \nabla $ is a linear connection. $g$ can be written as
  \ba
      g = \eta_{ab}e^a \otimes e^b
 \ea
where $\{ e^a \} $ are  orthonormal  co-frame $1-$forms and
$\eta_{ab} = \mbox{diag}(-1,+1)$. The orthonormal frame vectors
$\{ X_a \} $ dual to $\{ e^a \}$ are determined by the relations
$e^b(X_a)= \imath_a e^b = \delta^b_a $ where $\imath_a \equiv
\imath_{X_a}$ denote the interior product operators with respect
to the frame vectors $X_a$. We set the space-time orientation by
the choice $\epsilon_{01}=+1 $ or $*1 =e^0 \wedge e^1$ where $*$
is the Hodge map. Finally, the connection $\nabla$ is specified by
a set of $1$-forms $\{ {\Lambda^a}_b \}$. Cartan structure
equations below define the non-metricity, torsion and curvature
forms:
  \ba
    D\eta_{ab} := - (\Lambda_{ab}+\Lambda_{ba}) &=&  -2 Q_{ab} \; , \label{nonmet}\\
   De^a  := de^a + {\Lambda^a}_b \wedge e^b  &=&  T^a  \; , \label{tors}\\
   D {\Lambda^a}_b :=  d {\Lambda^a}_b + {\Lambda^a}_c \wedge
     {\Lambda^c}_b  & =& {R^a}_b \; . \label{curva}
 \ea
$\wedge$ , $d$ and  $D $ denote the exterior product, exterior and
covariant exterior derivatives, respectively. Bianchi identities
follow as integrability conditions:
 \ba
       DQ_{ab} &=& \frac{1}{2} ( R_{ab} +R_{ba}) \; , \label{bianc:0} \\
       DT^a    &=& {R^a}_b \wedge e^b \; , \label{bianc:1} \\
       D{R^a}_b &=& 0  \; . \label{bianc:2}
 \ea

\bigskip

\noindent The connection 1-forms can be uniquely decomposed as
 \ba
     \Lambda^{a}_{\; b} =  \omega^{a}_{\; b}
     + K^{a}_{\; b} + q^{a}_{\; b}  + Q^{a}_{\; b} \label{connect:dec}
 \ea
where $\omega^{a}_{\; b}$ are the Levi-Civita connection 1-forms
satisfying
 \ba
     \omega^{a}_{ \; b} \wedge e^b = - de^a , \label{LevCiv}
 \ea
$K^{a}_{\; b}$ are the contortion 1-forms satisfying
 \ba
   K^{a}_{\; b} \wedge e^b = T^a , \label{contort}
 \ea
and
 \ba
    q_{ab} = -( \imath_a Q_{bc} ) e^c + ( \imath_b Q_{ac}) e^c  \; . \label{q:ab}
 \ea

\bigskip

\noindent We also make use of the following identities
 \ba
    D*e_a &=& - Q \wedge *e_a + * (e_a \wedge e_b) \wedge T^b \label{ident:1}\\
    D*(e_a \wedge e_b) &=& -  Q \wedge *(e_a \wedge e_b)  \label{ident:2}
 \ea
where $ Q := \Lambda^a{}_a = \eta^{ab} \Lambda_{ab}$ is the trace
of the connection 1-forms.
\bigskip

\noindent We demand \cite{dereli} that as the metric  locally
re-scales as \ba g \rightarrow e^{2 \sigma} g \; ,\ea connection
should change according to \ba \Lambda^{a}_{\; b} \rightarrow
\Lambda^{a}_{\; b} - \eta^{a}_{\; b} d\sigma \; .  \ea Then we check
through the structure equations the following scaling rules: \ba T^a
\rightarrow e^{\sigma} T^a \quad , \quad R^{a}_{\; b} \rightarrow
R^{a}_{\; b} \quad , \quad Q_{ab} \rightarrow Q_{ab} - \eta_{ab}
d\sigma \; . \ea

\bigskip

\noindent In writing down gravity models it is convenient to obtain
first the irreducible decomposition of non-metricity tensor under
the 2D Lorentz group. We write
 \ba
   Q_{ab} = \underbrace{\overline{Q}_{ab}}_{trace-free \;\; part} + \underbrace{\frac{1}{2} \eta_{ab}Q}_{trace \;\;
   part} \; .
 \ea
Then
 \ba
     Q_{ab} = { }^{(1)}Q_{ab} + { }^{(2)}Q_{ab} + { }^{(3)}Q_{ab}
 \ea
in terms of
 \ba
    { }^{(1)}Q_{ab} &=& \frac{1}{2}[( \imath_a \Lambda) e_b + (\imath_b \Lambda) e_a - \eta_{ab} \Lambda ] \; , \\
    { }^{(2)}Q_{ab} &=& \frac{1}{2} \eta_{ab} Q  \; ,  \\
    { }^{(3)}Q_{ab}  &=& Q_{ab}- { }^{(1)}Q_{ab} - { }^{(2)}Q_{ab}
 \ea
where $ \Lambda := ( \imath_b { \overline{Q}^b}_a) e^a  \; .$ These
irreducible components satisfy
 \ba
   \eta_{ab} { }^{(1)}Q^{ab} = \eta_{ab} { }^{(3)}Q^{ab} = 0 \; ,
   \;
 \imath_a { }^{(3)}Q^{ab}  =0 \;, \;
  e_a \wedge { }^{(3)}Q^{ab} =0 \; .
 \ea
Thus they are orthogonal in the following sense:
 \ba
    { }^{(i)}Q^{ab} \wedge \; * { }^{(j)}Q_{ab} = \delta^{ij}
    N_{ij} \quad \quad (\mbox{\small no summation over {\it ij}})
 \ea
where $\delta^{ij}$ is the Kronecker symbol and $N_{ij}$ are certain
$2$-forms. We calculate
 \ba
 { }^{(1)}Q^{ab} \wedge \; * { }^{(1)}Q_{ab} &=& (\imath^a Q_{ac}) (\imath_b Q^{bc}) *1
                 + \frac{1}{4} Q \wedge * Q - (\imath_a Q) (\imath_b Q^{ab}) *1 \; , \nonumber \\
 { }^{(2)}Q^{ab} \wedge \; * { }^{(2)}Q_{ab} &=& \frac{1}{2}   Q \wedge * Q \;
 , \label{Q3Q3}
 \\
 { }^{(3)}Q^{ab} \wedge \; * { }^{(3)}Q_{ab} &=& Q^{ab} \wedge \;* Q_{ab}
     - { }^{(1)}Q^{ab} \wedge \; * { }^{(1)}Q_{ab} - { }^{(2)}Q^{ab} \wedge \; * { }^{(2)}Q_{ab}  \;
     . \nonumber
 \ea

 \section{ Symmetric Teleparallel Gravity}

We formulate STPG  by a variational principle  from an action
 \ba
   I = \int_M \left ( L + {R^a}_b {\rho_a}^b + T^a  \lambda_a  \right)
   \label{lagranj1}
 \ea
where $L$ is a Lagrangian density 2-form quadratic in the
non-metricity tensor, ${\rho_a}^b$ and $\lambda_a$ are  Lagrange
multiplier $0$-forms that impose the constraints
 \ba
   {R^a}_b = 0 \quad , \quad \quad T^a = 0 \; .
 \ea
The gravitational field equations are derived from (\ref{lagranj1})
by independent variations with respect to the connection $ \{
{\Lambda^a}_b \} $ and the orthonormal co-frame $\{ e^a \}$ (and the
Lagrange multipliers):
 \ba
     \lambda_a \wedge e^b + D{\rho_a}^b &=& - {\Sigma_a}^b  \; , \label{dLambda1}  \\
  D\lambda_a &=& - \tau_a  \label{dcoframe1}
 \ea
where $${\Sigma_a}^b = \frac{\partial L}{\partial {\Lambda^a}_b}
\quad , \quad \tau_a = \frac{\partial L}{\partial e^a} . $$ In
principle (\ref{dLambda1}) is  solved for the Lagrange multipliers
$\lambda_a$ and ${\rho_a}^b$ and substituted in (\ref{dcoframe1})
that governs the dynamics of the gravitational fields. It is
important to notice that $D \lambda_a$ rather than the Lagrange
multipliers themselves appear in  (\ref{dcoframe1}). Thus  we must
be calculating $D\lambda_a$ directly and that we can do by taking
the covariant exterior derivative of (\ref{dLambda1}):
 \ba
     D\lambda_a \wedge e^b = - D {\Sigma_a}^b \; .  \label{Dlambda}
 \ea
 We used in the above derivation,  the constraints
 \ba
 T^a = 0 \quad ,\quad
 D^2{\rho_a}^b = {R^b}_c \wedge {\rho_a}^c - {R^c}_a \wedge
 {\rho_c}^b =0 \label{D2Sab}
 \ea
where the covariant exterior derivative of a $(1,1)$-type tensor is
given by
 \ba
    D{\rho_a}^b = d {\rho_a}^b + {\Lambda^b}_c \wedge {\rho_a}^c
         - {\Lambda^c}_a \wedge {\rho_c}^b \; .
 \ea
Finally we arrive at the gravitational field equations
 \ba
     D{\Sigma_a}^b -  \tau_a \wedge e^b =0 \; .
 \ea
 In order to use symmetry arguments we lower an index:
 \ba
 D\Sigma_{ab} + 2 {Q^c}_b \wedge \Sigma_{ac} - \tau_a \wedge e_b
 =0 \; . \label{fieldeqn}
 \ea
The remaining set of field equations  are derived via the covariant
exterior derivative of (\ref{dcoframe1}):
 \ba
     D\tau_a =0  \label{Dtaua}
 \ea
where we used $D^2 \lambda_a = -{R^b}_a \lambda_b =0$. Note that our
system of equations (\ref{fieldeqn}) and (\ref{Dtaua}) is complete
in the sense that the total number of unknowns is 12 = 8
(connection) + 4 (co-frame), while the total number of  equations is
also 12 = 4 (vanishing curvature)+ 2 (vanishing torsion) + 4
(eqn.(\ref{fieldeqn})) + 2 (eqn.(\ref{Dtaua})).

\medskip

\noindent We now write down the most general Lagrangian density
2-form which is quadratic in the non-metricity tensor
\cite{obukhov1997}:
 \ba
   L =  \sum_{I=1}^3 k_I { }^{(I)}Q_{ab} \wedge * { }^{(I)}Q^{ab}
    + k_4 \left( { }^{(1)}Q_{ab} \wedge e^b \right) \wedge * \left( { }^{(2)}Q^{ac} \wedge e_c
    \right)   \label{Lagrange}
   \ea
where  $k_1, k_2, k_3, k_4$ are dimensionless coupling constants.
 Inserting
(\ref{Q3Q3}) in (\ref{Lagrange}), we simplify it  to
 \ba
   L =   c_1 Q_{ab} \wedge * Q^{ab}
      + c_2 (\imath_a Q^{ac}) (\imath^b Q_{bc}) *1 + c_3 Q \wedge * Q
       +c_4 (\imath_a Q) (\imath_b Q^{ab}) *1  \label{lagranj}
   \ea
in terms of new coupling constants
 \ba
 c_1 &=& k_3 ,\nonumber \\
 c_2 &=&  k_1 - k_3 \; ,\nonumber \\
 c_3 &=&  \frac{1}{4} k_1 + \frac{1}{2} k_2 - \frac{3}{4} k_3 + \frac{1}{4} k_4 \;, \nonumber \\
 c_4 &=& - k_1 + k_3 - \frac{1}{2} k_4 \; .
 \ea
One should note that for the choice
 \ba
   c_1 = -1/2 \; , \;\; c_2 = 1 \; , \;\; c_3 = 1/2 \; , \;\;c_4 = -1 , \label{EHvalues1}
 \ea
the Lagrangian density (\ref{lagranj}) is an exact form. This is
proven in the appendix. Furthermore the Lagrangian density
(\ref{lagranj}) is scale invariant if and only if \ba
   c_1 + 2 c_3 = 0 \; , \;c_2 = c_4 = 0 \; \; . \label{scaleinv}
 \ea
The model has no dynamics for the above two cases. We expect
non-trivial solutions in all the other cases.

The variation of the action  with (\ref{lagranj}) is long but
straightforward.  We list below the  contributions to the
variational field equations (\ref{fieldeqn})-(\ref{Dtaua}) coming
from each term in (\ref{lagranj}):
 \ba
   {\Sigma_a}^b = \sum_{i=0}^{4} c_i \; { }^{(i)}{\Sigma_a}^b \quad ,
   \quad \quad \quad \quad
  \tau_a = \sum_{i=0}^{4} c_i \; { }^{(i)}\tau_a
 \ea
where
 \ba
  { }^{(1)} \Sigma_{ab} &=& 2*Q_{ab} \; , \nonumber \\
  { }^{(2)} \Sigma_{ab} &=& \imath^c Q_{ac} *e_b + \imath^c Q_{bc} *e_a \; , \nonumber \\
  { }^{(3)} \Sigma_{ab} &=& 2 \eta_{ab} *Q \; ,\nonumber \\
  { }^{(4)} \Sigma_{ab} &=& \frac{1}{2} (\imath_a Q) *e_b +  \frac{1}{2} (\imath_b Q) *e_a
                                            + \eta_{ab}  (\imath_c Q^{cd}) *e_d \;
                                            , \label{SigmaQ}
  \ea
  and
  \ba
  { }^{(1)}\tau_a  &=& - (\imath_a Q^{bc}) \wedge *Q_{bc} - Q^{bc} \wedge (\imath_a *Q_{bc}) \; , \nonumber \\
  { }^{(2)}\tau_a  &=& (\imath_b Q^{bc}) (\imath^d Q_{cd}) *e_a
                                - 2 (\imath_a Q^{bd}) (\imath^c Q_{cd}) *e_b \; , \nonumber \\
  { }^{(3)}\tau_a  &=& -(\imath_a Q) \wedge *Q - Q \wedge (\imath_a *Q) \; ,\nonumber \\
  { }^{(4)}\tau_a  &=& (\imath_b Q) (\imath_c Q^{bc}) *e_a - (\imath_a Q) (\imath_c Q^{bc}) *e_b
                            - (\imath_b Q) (\imath_a Q^{bc}) *e_c \;
                            . \label{tauQ}
 \ea

\section{Solutions }

The Minkowski metric with vanishing non-metricity; \ba g = - dT^2 +
dR^2 \quad , \quad Q_{ab} = 0 \ea is a trivial solution of our
system. We first consider a class of conformally flat solutions
determined by  metric
 \ba
  g = e^{2 \psi(T,R)} ( - dT^2 + dR^2 )
   \ea
and the non-metricity 1-forms
 \ba
 Q_{ab} = -\eta_{ab} d\psi (T,R)  \label{Q_ab}
 \ea
that is generated by a conformal re-scaling of the flat solution
with an arbitrary function $\psi(T,R)$. This configuration satisfies
the constraints ${R^a}_b(\Lambda) = 0$ and $T^a(\Lambda) = 0$. The
non-trivial equations come from the trace and the symmetric parts of
the field equation (\ref{fieldeqn}), and the zeroth and the first
components of the field equation (\ref{Dtaua}). They are,
 \ba
   \kappa (\psi_{TT} - \psi_{RR}) =0 \\
   \chi \psi_{TR} + \kappa \psi_T \psi_R =0 \\
   2\kappa \psi_T (\psi_{TT} - \psi_{RR}) =0 \\
   2\kappa \psi_R (\psi_{TT} - \psi_{RR}) =0
 \ea
where
 \ba
  \kappa = 2c_1 + c_2 + 4c_3 +2 c_4 \quad , \quad
  \chi = c_2 + c_4 \; .
 \ea
Here the lower indices $T$ and $R$ denote derivatives with respect
to coordinates $T$ and $R$, respectively.  It is checked
$\kappa=\chi =0$ for the values (\ref{EHvalues1}) and
(\ref{scaleinv}) which means that any choice of $\psi(T,R)$ gives
a solution. Otherwise, the general solution is given by
 \ba
     \psi (T,R) &=& \frac{\chi}{\kappa} \ln \left\{ \cosh \left[ \frac{a_1 \sqrt{\kappa}}{\chi}
                    (R+T+a_2)\right] \right\} \nonumber \\
     & & + \frac{\chi}{\kappa} \ln \left\{ \cosh \left[ \frac{a_1 \sqrt{\kappa}}{\chi}
              (R-T+a_3)\right] \right\} +a_4 \nonumber \\
        &=& \frac{\chi}{\kappa} \ln \left\{ \cosh \left[ \frac{a_1 \sqrt{\kappa}}{\chi} (2R+a_2+a_3) \right] \right. \nonumber \\
       & & \left. + \cosh \left[ \frac{a_1 \sqrt{\kappa}}{\chi} (2T+a_2-a_3) \right]
       \right\} - \frac{\chi}{\kappa} \ln 2 +a_4
 \ea
where $a_1,a_2,a_3,a_4$ are integration constants.
 We also give
 \ba
  d\psi = \frac{a_1}{\sqrt{\kappa}} (dR + dT)
                    \mbox{tanh} \left[ \frac{a_1 \sqrt{\kappa}}{\chi}(R+T+a_2) \right] \nonumber \\
   + \frac{a_1}{\sqrt{\kappa}} (dR - dT) \mbox{tanh}\left[ \frac{a_1 \sqrt{\kappa}}{\chi}(R-T+a_3)
   \right] \; .
 \ea

\bigskip

We wish now to discuss two other classes of solutions that may or
may not be included in the class of conformally flat solutions
above, depending on the parameter values.

 \subsection{Static solutions}

Suppose $\psi = \psi(R)$ only. Then consider a coordinate
transformation $ (T,R) \rightarrow (t,r)$ such that
$$
t = T \; , \; r = \int e^{2 \psi(R)} dR \;
$$
and let
$$
\psi(R) = ln |f(r)| \; .
$$
In the new coordinate chart our orthonormal co-frame reads
 \ba
 e^0 = f(r)dt \quad , \quad e^1 = dr/f(r)  \quad .
 \ea
We make the following ansatz independently for the non-metricity
1-forms \ba
    Q_{00} = f_r e^1 \quad , \quad Q_{11} = h(r)
     e^1 \quad , \quad Q_{01} = Q_{10} = 0 \;
     \ea
where $h(r)$ is arbitrary.  Here lower index $r$ denotes
derivative with respect to $r$. This configuration satisfies the
constraints ${R^a}_b(\Lambda) =0$ and $T^a(\Lambda) =0$.  The only
non-trivial contribution to the field equation (\ref{fieldeqn})
comes from its trace: $d{\Sigma^a}_a =0$, and the first component
of the field equation (\ref{Dtaua}): $D\tau_1 =0$. The field
equations reduce to
  \ba
 \alpha \left( fh \right)_r - \beta \left( ff_r \right)_r &=& 0 \label{eqn1} \\
  \gamma (f_r)^2 h + \mu (h^3 -2fhh_r)- \nu [(f_r)^3 +  2ff_rf_{rr}] & & \nonumber \\
  - \xi h^2 f_r + \eta (hff_{rr} + ff_rh_r)&=& 0 \label{eqn2}
 \ea
where
 \ba
    \alpha &=& 2c_1 +2c_2 +4c_3 + 3c_4 \nonumber \\
    \beta &=& 2c_1 + 4c_3 + c_4 \nonumber \\
    \gamma &=& c_1 + 3c_3 + c_4 \nonumber \\
   \mu &=& c_1 + c_2 + c_3 + c_4 \nonumber \\
   \nu &=& c_1 + c_3 \nonumber \\
   \xi &=& c_1 + c_2 + 3c_3 + 2c_4 \nonumber \\
   \eta &=& 2c_3 + c_4 \; . \label{alpha-sigma}
 \ea
It is checked that all these constants are zero for the values
(\ref{EHvalues1}) which means that any choice of $f(r)$ and $h(r)$
gives a solution. For a non-trivial solution we first solve for
$h$ from Eqn.(\ref{eqn1}):
 \ba
     h= \frac{\beta}{\alpha} f_r \; .
 \ea
Substituting the above expression into (\ref{eqn2}) yields
 \ba
     A ff_{rr} + B (f_r)^2 =0
 \ea
where
 \ba
    A &=& 2\eta \beta \alpha^2 - 2\mu \alpha \beta^2 - 2 \nu \alpha^3 \nonumber \\
    B &=& \gamma \beta \alpha^2 + \mu \beta^3 - \nu \alpha^3 - \xi \alpha \beta^2 \;
    . \label{AB}
 \ea
Thus a solution  is found as
 \ba
     f(r) = (a_1 + a_2 r)^{\frac{A}{A+B}} \label{solut}
 \ea
where $a_1, a_2$ are integration constants. We note that if
$\beta=-\alpha$, we can write the full connection 1-form as an exact
form $\Lambda_{ab} = - \eta_{ab} d(\ln f (r))$.

We want to remark briefly on the singularity structure of the above
solution. This is important for its physical interpretation. Within
the simplest approach to singularities we examine scalars formed out
of the main ingredients of space-time; e.g. $\mathcal{R}(\omega) =
\imath_b \imath_a R^{ab}(\omega)$ in GR. In our case this amounts to
looking at the quadratic non-metricity invariants such as $Q_{ab}
\wedge *Q^{ab}$. Thus, from the Riemannian point of view we conclude
that, since
 \ba
     \mathcal{R}(\omega) = \frac{2{a_2}^2A(B-A)}{(A+B)^2} \frac{1}{(a_1 + a_2 r)^{\frac{2B}{A+B}}}
 \ea
the surface $r=-a_1/a_2$ is an essential singularity (provided
$\frac{2B}{A+B} > 0$). On the other hand from the symmetric
teleparallel point of view we find
 \ba
    Q_{ab} \wedge *Q^{ab} = \left( 1+ \frac{\beta^2}{\alpha^2} \right)
      \left( \frac{a_2 A}{A+B} \right)^2  \frac{1}{(a_1 + a_2 r)^{\frac{2B}{A+B}}} *1 \; .
 \ea
We see that the singularity structure of the above solution of STPG
is basically similar to that of GR
\cite{adak2005},\cite{obukhov2003}.

 \subsection{Cosmological solutions}

Suppose $\psi = \psi(T)$ only. Then consider a coordinate
transformation $ (T,R) \rightarrow (t,r)$ such that
$$
t = \int e^{ \psi(T)} dT \; , \;  r = R \; .
$$
and let
$$
\psi(T) = ln |{\mathcal{R}}(t)| \; .
$$
In the new coordinate chart our orthonormal co-frame reads \ba e^0=
dt \quad , \quad e^1 = {\mathcal{R}}(t) dr  \quad .
 \ea
We start with an independent ansatz for the non-metricity 1-forms
 \ba
     Q_{00} = \mathcal{S}(t) e^0 \quad , \quad  Q_{11} = -
     \frac{\mathcal{R}_t}{\mathcal{R}} e^0 \quad , \quad  Q_{01} = Q_{10} = 0
 \ea
where lower index $t$ denotes derivative with respect to $t$. For
this configuration, the only non-trivial part of (\ref{fieldeqn}) is
again its trace;  and the zeroth component of (\ref{Dtaua}) which
reads
 \ba
    \alpha (\mathcal{RS})_t + \beta \mathcal{R}_{tt} &=& 0 \\
    \xi  \mathcal{R}^2 \mathcal{S}^2 \mathcal{R}_t + \mu (\mathcal{R}^3 \mathcal{S}^3 + 2 \mathcal{R}^3 \mathcal{S S}_t) \nonumber & & \\
    + \nu [ 2\mathcal{R R}_t \mathcal{R}_{tt} - (\mathcal{R}_t)^3 + \mathcal{RS} (\mathcal{R}_t)^2 ] + \eta (\mathcal{SR R}_{tt} +
     \mathcal{R}^2 \mathcal{R}_t \mathcal{S}_t) &=& 0
 \ea
where all constants are defined in (\ref{alpha-sigma}). The solution
to these two equations is given by
 \ba
    \mathcal{ S} = - \frac{\beta}{\alpha} \frac{\mathcal{R}_t}{\mathcal{R}} \quad , \quad
    \mathcal{ R}(t) = (b_1 + b_2 t)^n
 \ea
where $b_1, b_2$ are integration constants and
 \ba
     n= \frac{A}{A+C} \; .
 \ea
Here $A$ is given by (\ref{AB}) and
 \ba
    C = \mu \beta^3 + 2 \mu \alpha \beta^2 + \nu \alpha^3 + \nu
    \beta \alpha^2 - \eta \beta \alpha^2 - \alpha \xi \beta^2 \;  .
 \ea
We determine a particular solution by setting $b_1 = 1$ and $b_2 = H
/ n$, and using the fact
 \ba
     \lim_{n \to \infty} \left( 1+ \frac{H}{n}t \right)^n = e^{Ht}
     \; ,
 \ea
in the limit $C \to -A$. Then the metric takes a special form that
satisfies the perfect cosmological principle:
 \ba
 g  = -dt^2 + e^{2Ht} dr^2 .
 \ea
 Now a  co-ordinate transformation $\tan{r}= \rho$  yields the closed Robertson-Walker metric
 \ba
    g  = -dt^2 + e^{2Ht}\frac{d\rho^2}{(1+ \rho^2)^2} \; ,
\ea while another  co-ordinate transformation $\tanh{r}= \rho$
yields the open Robertson-Walker metric
 \ba
    g  = -dt^2 + e^{2Ht}\frac{d\rho^2}{(1 - \rho^2)^2} \; .
\ea We again note that if we set $\beta=-\alpha$, we can write the
full connection 1-form an exact form $\Lambda_{ab} = - \eta_{ab}
d(\ln |\mathcal{R }(t)|)$.

\section{Conclusion}

Theories of symmetric teleparallel gravity (STPG)  in which the
non-metricity tensor is dynamical are not easy to interpret
\cite{dereli}. However, they may play a role in certain
astrophysical contexts. In addition to ordinary matter that
couples gravitationally through its mass, we may think of dark
matter in the universe that couples to a new kind of gravitational
charge, called the Weyl charge \cite{tucker}, through the
non-metricity tensor. The search for classical solutions is
important to understand any observational effects that would be
induced by such couplings. In this paper we investigated a
symmetric teleparallel gravity model in two-dimensions. Firstly
after writing scale transformation rules in non-Riemannian
geometries, we discussed the decomposition of the full connection
and orthogonal, irreducible decomposition of the non-metricity
tensor under the 2D Lorentz group. Then  we wrote the most general
four-parameter Lagrangian density $2$-form that is quadratic in
the non-metricity tensor. For a specific set of values of the
coupling constants the Lagrangian is an exact form and thus leads
to a topological invariant.  Besides, for some other special
values of the coupling constants, the Lagrangian is scale
invariant which means that the model does not have dynamics. We
derived the variational field equations and found a class of
conformally flat solutions. We also gave two other classes of
solutions, one of which may be related to static, spherically
symmetric solutions of 4D gravity while the other one takes the
familiar form of an inflationary cosmological metric in 4D.


 \newpage

 \section{Appendix}

We prove that for the following values of the coupling constants
 \ba
   c_1 = -1/2 \; , \;\; c_2 = 1 \; , \;\; c_3 = 1/2 \; , \;\;c_4 = -1 , \label{EHvalues}
 \ea
Lagrangian (\ref{lagranj}) is an exact
 form:
 \ba
     L = d(\omega + q) \label{EH-equv}
 \ea
where $\omega_{ab} = \epsilon_{ab} \omega $ and $q_{ab} =
\epsilon_{ab}q$. This means that the Lagrangian (\ref{lagranj}) with
the values (\ref{EHvalues}) do not yield dynamical field equations.
 We decompose the
non-Riemannian curvature scalar by substituting the decomposition of
the full connection ( with ${K^a}_b =0$ ), into the non-Riemannian
curvature 2-form ${R^a}_b (\Lambda)$:
 \ba
  \mathcal{R} (\Lambda) *1 &=& \frac{1}{2} {R^a}_b (\Lambda) \wedge *{e_a}^b  \nonumber \\
   &=& \frac{1}{2} \epsilon_{ab} R^{ab} (\omega) + \frac{1}{2} d(\epsilon_{ab} q^{ab})
       + \frac{1}{2} \epsilon_{ab} Q^a{}_c \wedge  Q^{bc}  \; .
 \ea
In STPG ${R^a}_b (\Lambda)=0$ so that $\mathcal{R} (\Lambda) = 0$
and we have $ R^{ab}(\omega) =  d\omega^{ab}$ in two-dimensions.
Therefore
 \ba
  {\epsilon_a}^b Q^{ac} \wedge Q_{cb} =  - \epsilon_{ab}
 ( d\omega^{ab} +  dq^{ab}) \; . \label{EH-1}
 \ea
Here we write the left hand side in terms of the volume form as
 \ba
     {\epsilon_a}^b Q^{ac} \wedge Q_{cb} = (\imath_a Q^{ac}) (\imath^b Q_{bc})
     *1 - (\imath^a Q^{bc}) (\imath_c Q_{ab}) *1
 \ea
and calculate the second term of the right hand side of this
expression as follows:
 \ba
     0 &=& {\epsilon_a}^b q^{ac} \wedge q_{cb} \nonumber \\
      &=&  Q_{ab} \wedge * Q^{ab} - (\imath_a Q^{ac}) (\imath^b Q_{bc}) *1 - Q \wedge *
      Q \nonumber \\
    & &   + 2 (\imath_a Q) (\imath_b Q^{ab}) *1  - (\imath^a Q^{bc}) (\imath_c Q_{ab}) *1 \; .
 \ea
Finally we insert these results into (\ref{EH-1}) together with
$\omega_{ab} = \epsilon_{ab} d\omega$ and $q_{ab} = \epsilon_{ab} q$
to get
 \ba
   d(\omega + q) = - \frac{1}{2} Q_{ab} \wedge * Q^{ab}
      + (\imath_a Q^{ac}) (\imath^b Q_{bc}) *1 + \frac{1}{2} Q \wedge * Q \nonumber \\
        - (\imath_a Q) (\imath_b Q^{ab}) *1
 \ea
which leads to  (\ref{EHvalues}) and (\ref{EH-equv}).

\end{document}